\documentclass[a4paper,fleqn]{cas-dc}

\usepackage[switch]{lineno}
\usepackage{ulem}
\usepackage{gensymb}
\usepackage[table]{xcolor}
\usepackage[authoryear]{natbib}
\usepackage{amssymb}
\usepackage{amsmath}
\usepackage{booktabs} 
\usepackage{multirow}
\usepackage{makecell}
\usepackage{url}
\usepackage{comment}
\usepackage[section]{placeins} 
\usepackage{url}
\usepackage{stfloats}  
\usepackage{siunitx}
\usepackage{threeparttable}
\newcommand{\sym}[1]{\rlap{#1}}

\def\tsc#1{\csdef{#1}{\textsc{\lowercase{#1}}\xspace}}
\tsc{WGM}
\tsc{QE}

\begin{document}
\let\WriteBookmarks\relax
\def\floatpagepagefraction{1}
\def\textpagefraction{.001}

\shorttitle{ORBIT}    

\shortauthors{Yang et al.}  

\title [mode = title]{Orientation-Robust Latent Motion Trajectory Learning for Annotation-Free Cardiac Phase Detection in Fetal Echocardiography}  



%

\author[1]{Yingyu Yang}

\cormark[1]


\ead{yingyu.yang@eng.ox.ac.uk}


\author[1]{Qianye Yang}
\author[1]{Can Peng}
\author[2]{Elena D'Alberti}
\author[2]{Olga Patey}
\author[2]{Aris T. Papageorghiou}
\author[1]{J. Alison Noble}
\affiliation[1]{organization={Institute of Biomedical Engineering, Department of Engineering Science, University of Oxford}, city={Oxford},country={United Kingdom}}

\affiliation[2]{organization={Nuffield Department of Women’s and Reproductive Health, University of Oxford}, city={Oxford},country={United Kingdom}}

\cortext[1]{Corresponding author}



\begin{abstract}
Fetal echocardiography is essential for detecting congenital heart disease (CHD), facilitating pregnancy management, optimized delivery planning, and timely postnatal interventions. Among standard imaging planes, the four-chamber view (4CV) provides important information for CHD diagnosis, where clinicians carefully inspect the end-diastolic (ED) and end-systolic (ES) phases to evaluate cardiac structure and motion. Automated detection of these cardiac phases is thus a critical component towards fully automated CHD analysis. However, existing approaches typically rely on manual annotation of ED/ES frames, which is labour-intensive and time-consuming.
We present ORBIT (Orientation-Robust Beat Inference from Trajectories), a self-supervised framework that identifies cardiac phases without manual annotations under various fetal heart orientation. ORBIT employs registration as self-supervision task and learns a latent motion trajectory of cardiac deformation, whose turning points capture transitions between cardiac relaxation and contraction, enabling accurate and orientation-robust localization of ED and ES frames across diverse fetal positions. 
Trained exclusively on normal fetal echocardiography videos, ORBIT achieves consistent performance on both normal (mean absolute error 1.9 frames for ED and 1.6 for ES) and CHD cases (mean absolute error 2.4 frames for ED and 2.1 for ES), outperforming existing annotation-free approaches constrained by fixed orientation assumptions.
These results highlight the potential of ORBIT to facilitate robust cardiac phase detection directly from 4CV fetal echocardiography. 
\end{abstract}





\maketitle

\section{Introduction}
\label{intro}

Congenital heart disease (CHD) is one of the most common birth defects in the fetal population and remains a major global burden for child health \citep{zimmerman2020global}. 
Prenatal screening for CHD plays a critical role in reducing perinatal mortality, optimizing postnatal management, and supporting pregnancy planning \citep{hunter2014prenatal}. Fetal echocardiography is an effective modality for detecting and evaluating CHD \citep{donofrio2014diagnosis}. However, performing fetal echocardiography and accurately identifying CHD require intensive training in free-hand ultrasound and specialized expertise in fetal cardiology, posing challenges for early CHD detection, particularly in low- and middle-income countries \citep{zuhlke2019congenital}. Therefore, improving the workflow for CHD detection in fetal echocardiography is of great importance \citep{patey2025prenatal}. Artificial intelligence (AI) has shown strong potential in assisting fetal echocardiography analysis, including improving CHD detection rates \citep{bridge2017automated,maraci2017framework,arnaout2021ensemble,sarker2023comformer,d2025artificial,mishra2025selfsupervised}.
A fundamental step in echocardiography analysis is the identification of the end-diastole (ED) and end-systole (ES) frames, as these two key cardiac phases are critical for functional assessment and cardiac biometric measurements in both adult \citep{lang2015recommendations} and fetal echocardiography \citep{crispi2013ultrasound,moon2023guidelines,carvalho2023isuog}. In practice, ED and ES correspond to the closure of the mitral and aortic valves, respectively \citep{mada2015define}. When these valves are not clearly visible, surrogate parameters such as maximal and minimal ventricular volumes for ED and ES, respectively, can be used to define these phases \citep{mada2015define}.

\begin{figure}[]
    \centering
    \includegraphics[width=\columnwidth]{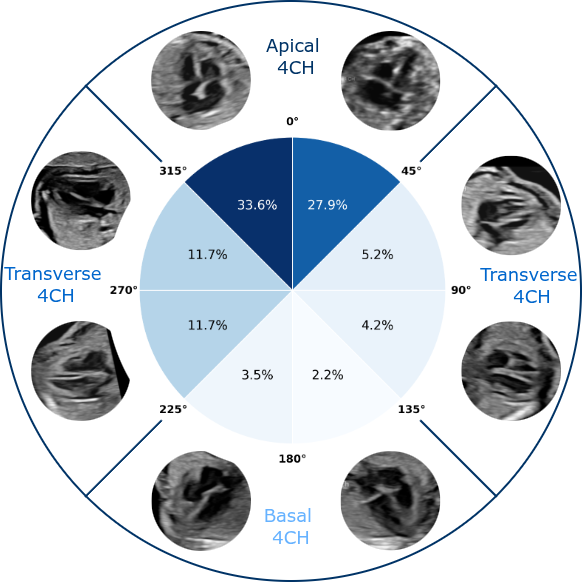}
    \caption{\textbf{Orientation distribution in the training set.} Fetal 4CV echocardiography exhibits greater orientation variability, primarily due to differences in fetal position and examination angle.}
    \label{fig_ori_count_trainvalid}
\end{figure}

Numerous automatic algorithms have been developed to detect ED/ES frames in adult \citep{dezaki2018cardiac,lane2021multibeat,li2023semi} and fetal echocardiography \citep{lee2020automatic,pu2021fetal,pu2024hfsccd,balaji2024optimized,lu2025optical}. However, these methods rely on training data with manual annotations, such as ED/ES indices or segmentation labels, limiting their applicability in annotation-scarce settings. Although several unsupervised approaches have been proposed for adult echocardiography \citep{gifani2010automatic,shalbaf2015echocardiography,ddsb,sjoerdsma2025spatio}, direct adaptation to fetal echocardiography remains challenging due to non-standardized imaging windows, variable fetal heart orientations, and potential fetal motion. Unlike adult 4-chamber views (4CV), typically acquired from a fixed imaging angle, fetal 4CV echocardiography exhibits substantial orientation variability due to differences in fetal position. As illustrated in Figure \ref{fig_ori_count_trainvalid}, approximately 60\% of our training data are apical 4CV, while the remaining 40\% consist of transverse or basal views. These challenges underscore the need for a unified model that can handle fetal heart orientation variability in fetal cardiac phase detection.

To overcome the challenges of orientation variability and the availability of annotations in fetal echocardiography, we propose ORBIT (Orientation-Robust Beat Inference from Trajectories), a novel self-supervised framework for fetal cardiac phase detection. ORBIT is annotation-free, orientation-robust, and effective across both normal and CHD cases.

The main contributions of this work are as follows:
\begin{itemize}
\item Annotation-free orientation-robust phase detection: We introduce a self-supervised method for identifying ED and ES frames in fetal 4CV echocardiography. The model learns a low-dimensional latent motion trajectory (1D/2D) that captures cardiac relaxation and contraction directly from video data without any manual labels and is shown to generalize across diverse fetal heart orientations.
\item Comprehensive evaluation of robustness: The model is trained solely on normal fetal echocardiography videos and evaluated across different orientation angles, CHD conditions, and preprocessing (cropping) strategies to assess both in- and out-of-distribution robustness.
\end{itemize}

This study builds upon our earlier work~\citep{yang2025latent}, which introduced an annotation-free framework for cardiac phase detection in adult and fetal apical 4CV views. In this extension, we make several key advances:
\begin{enumerate}
    \item We replace reconstruction-based self-supervision with a registration-based formulation, shifting the modelling target from echocardiographic appearance reconstruction to cardiac deformation estimation. The latent decomposition framework is used to encode the estimated temporal deformation into a compact, low-dimensional trajectory. Consequently, the latent trajectory represents cardiac motion dynamics rather than image appearance variation, improving robustness to variations in fetal heart orientation and enabling more accurate identification of cardiac phases;
    \item We double the number of normal videos used for in-distribution evaluation (88 vs. 44 in the previous work);
    \item We conduct a detailed evaluation on 156 CHD videos to assess out-of-distribution robustness; and
    \item We compare manual and tool-based cropping strategies to examine the preprocessing influence on self-supervised cardiac phase detection.
\end{enumerate}

The remainder of this paper is organized as follows. Section \ref{work} reviews related work on supervised and unsupervised cardiac phase detection and low-dimensional motion representation. Section \ref{method} presents the ORBIT framework. Section \ref{exp} describes the dataset, implementation details, and experimental setup. Section \ref{result} reports results on in-distribution and orientation performance as well as robustness to domain shifts (normal vs. CHD, manual vs. tool-based cropping). Finally, Section \ref{dis} concludes the paper.

\section{Related work}
\label{work}
\subsection{Supervised cardiac phase detection in adult and fetal echocardiography}
Supervised cardiac phase detection in 4CV echocardiography requires a large amount of annotated data, either through direct or indirect supervision. In adult echocardiography, direct approaches typically involve regressing the ED/ES frames directly from echocardiography inputs using neural networks, such as CNN-RNN frameworks~\citep{dezaki2018cardiac,lee2020automatic} or Vision Transformers~\citep{reynaud2021ultrasound}, to estimate temporal ED/ES probabilities. Indirect approaches infer ED/ES timing from intermediate temporal predictions, such as left ventricular segmentation masks~\citep{zeng2023maef} or derived volume curves~\citep{li2023semi}.  

In fetal echocardiography, a hybrid object detection framework has been proposed to jointly perform standard plane recognition, cardiac cycle localization, and abnormal structure detection in both apical and basal four-chamber views~\citep{pu2024hfsccd}. An optimized pyramidal convolutional network with shuffle-attention has been developed for improved ED and ES detection, with separate models trained for apical, basal, and parasternal 4CV~\citep{balaji2024optimized}. More recently, long-term temporal dependencies and short-term motion cues have been jointly modelled by integrating optical flow with a Mamba U-Net architecture to regress frame-wise ED/ES probabilities in fetal echocardiography videos~\citep{lu2025optical}.

Although promising, supervised methods are fundamentally constrained by the amount of annotated data available for training. This limitation hinders their applicability in annotation-scarce settings, especially in fetal echocardiography, where large quantities of unlabelled video data are available but expert annotations are costly and time-consuming to obtain.

\subsection{Unsupervised cardiac phase detection in adult and fetal echocardiography}
Unsupervised methods for ED/ES detection aim to exploit the intrinsic periodicity of cardiac motion in 4CV echocardiography, without the help of explicit labels. 

Dimension-reduction techniques, such as locally linear embedding (LLE), have been used to project echocardiography videos onto a low-dimensional manifold, where cardiac phase transitions can be inferred through density analysis~\citep{gifani2010automatic,shalbaf2015echocardiography}. However, these case-specific embeddings often lack interpretability and consistency when applied to larger datasets. 
Applying singular value decomposition (SVD) to spatio-temporal matrices derived from adult echocardiography videos has shown that the leading temporal singular vector can delineate cardiac phase transitions, although this analysis was demonstrated on a relatively small dataset of 20 cases~\citep{sjoerdsma2025spatio}.
Considering the inherently periodic nature of cardiac motion, circular latent representations have been explored using autoencoder-based frameworks, in which the cardiac cycle is constrained to follow a cyclic trajectory in latent space~\citep{laumer2020deepheartbeat}. While the resulting latent representations exhibit semantic alignment with ED and ES phases, these approaches do not explicitly predict their temporal locations.
Related training-free unsupervised methods have modelled left ventricular expansion–contraction dynamics to identify ED/ES frames in adult echocardiography videos; although computationally efficient, such approaches typically require careful parameter tuning when transferred to new datasets~\citep{ddsb}.

Despite these advances, applications of unsupervised cardiac phase detection in fetal echocardiography remain limited. Our earlier work~\citep{yang2025latent} proposed a self-supervised reconstruction-based framework called LMP (Latent Motion Profiling) with an interpretable latent subspace for studying cardiac phase dynamics, achieving promising ED/ES detection accuracy in both adult and fetal echocardiography. However, that method was restricted to apical 4CV views only and required manual orientation correction for other 4CV orientations, limiting its flexibility in handling fetal echocardiography data with diverse orientations. While orientation variability could, in principle, be mitigated by predicting heart orientation explicitly~\citep{huang2017temporal}, our proposed method eliminates this requirement altogether. ORBIT does not rely on any prior orientation information, making it substantially more versatile and scalable for minimally annotated datasets.

\subsection{Low-dimensional representation in cardiac motion modelling}
Cardiac phase is closely linked to cardiac motion. However, cardiac motion is inherently high-dimensional and highly non-linear. Representing it in a low-dimensional space enables more efficient modelling and analysis. Approaches such as barycentric subspace projection~\citep{rohe2018low} and polyaffine motion approximation~\citep{mcleod2015spatio,yang2023unsupervised} have been proposed to capture complex cardiac dynamics while reducing computational complexity. Latent probabilistic deep learning models have also been employed to learn conditional motion from end-diastole to the entire cardiac cycle, facilitating scalability to larger datasets~\citep{krebs2019probabilistic}.

Inspired by these non-linear low-dimensional motion modelling strategies, we design our framework to learn a latent subspace representation from the temporal deformation field across the entire input video, enabling efficient and generalizable characterization of cardiac motion.

\section{Methodology}
\label{method}

\begin{figure*}[h]
    \centering
    \includegraphics[width=\textwidth]{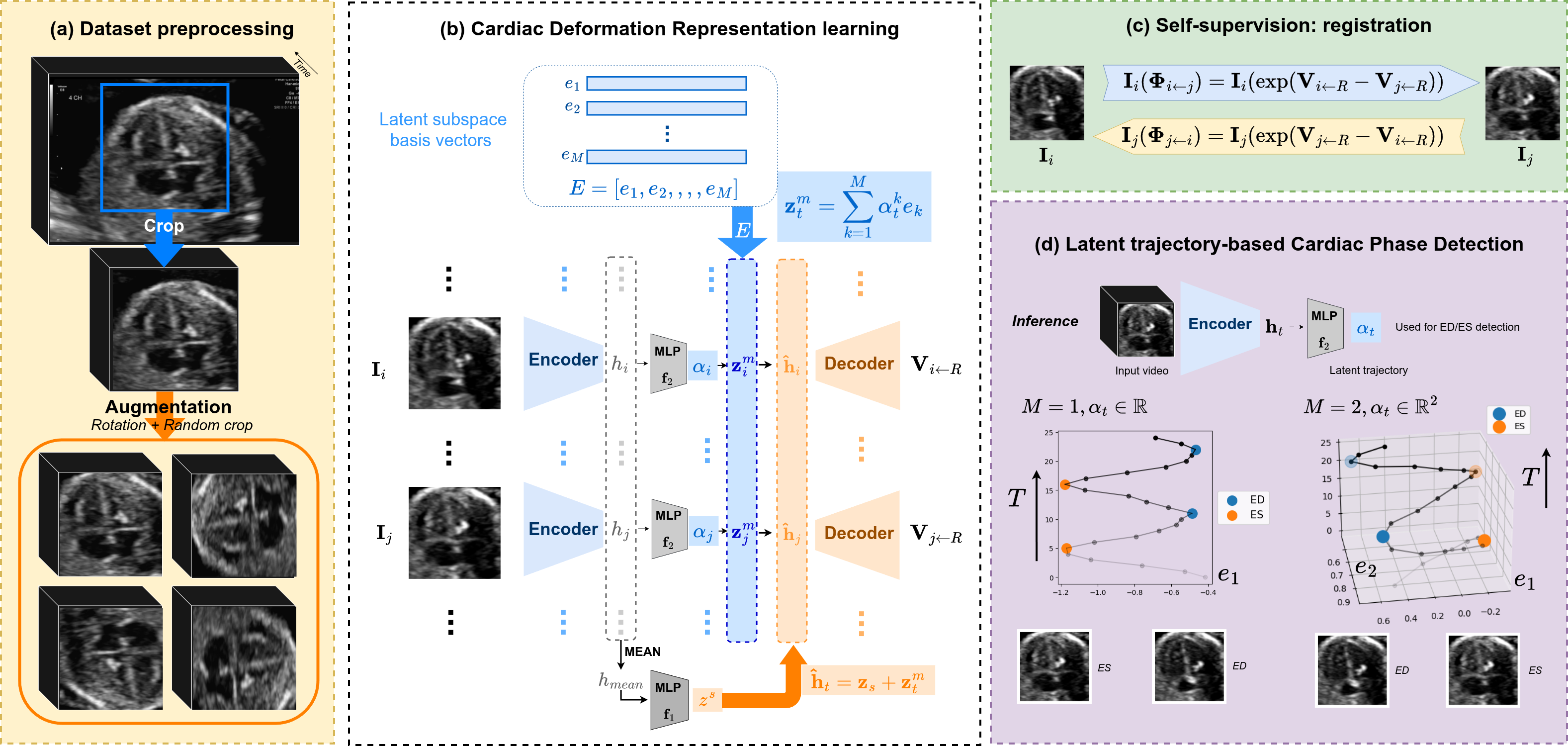}
    \caption{\textbf{Proposed ORBIT model for cardiac phase detection.} (a) Video preprocessing with heart area cropping and rotation augmentation (Section \ref{exp}).(b) Latent decomposition for cardiac deformation representation learning (Section \ref{method_rep}) and (c) its self-supervision aim via registration (Section \ref{method_aim}). (d) Annotation-free cardiac phase detection from latent trajectory with latent subspace dimension $M=1$ and $M=2$  (Section \ref{method_traj}).}
    \label{fig_svf_method}
\end{figure*}

Building on the insight that cardiac relaxation and contraction represent a global abstraction of highly complex heart motion, we hypothesize that a low-dimensional representation can capture cardiac phase information from high-dimensional deformation patterns. Therefore, we aim to learn the underlying deformation dynamics within an echocardiography video and, through latent space decomposition, characterize cardiac phase transitions along a low-dimensional latent trajectory. In the following, we describe how to learn this latent trajectory via a self-supervised registration task and how cardiac phase can be consistently aligned with the resulting explainable trajectory, enabling robust ED/ES detection across fetal echocardiography videos with varying orientations.

\subsection{Reference-Free Cardiac Deformation Estimation}
\label{method_aim}
Traditional cardiac deformation estimation methods typically require a reference frame, commonly the ED frame, to track motion relative to that frame~\citep{mcleod2015spatio,rohe2018low,krebs2019probabilistic,yang2023unsupervised}. However, in our setting, the ED frame is unknown because it is the target of our detection task.  
To avoid relying on a manually specified reference frame, we adopt the abstract reference frame formulation introduced in prior self-supervised motion representation works \citep{siarohin2019first,wang2024lia}. In this formulation, each observed frame is related to an abstract reference frame, which serves as an intermediate coordinate system for estimating deformation between arbitrary frame pairs.
Given a frame $\mathbf{I}_t : \mathbb{R}^2 \rightarrow \mathbb{R}$ from an echocardiography video, we use an autoencoder to estimate a deformation field $\mathbf{\Phi}_{t\leftarrow R}: \mathbb{R}^2 \rightarrow \mathbb{R}^2$ that warps the current frame to the abstract reference frame $\mathrm{R}$: 
\begin{equation}
    \mathbf{\hat{I}_t} = \mathbf{I}_t   \circ \mathbf{\Phi}_{t\leftarrow R} \approx \mathrm{R}
\end{equation}

Unlike the original formulation~\citep{siarohin2019first,wang2024lia}, we parameterize the transformation to the abstract reference frame using a stationary velocity field (SVF) $\mathbf{V}_{t \leftarrow R}$, such that $\Phi_{t \leftarrow R} = \exp(\mathbf{V}_{t \leftarrow R})$~\citep{arsigny2006log}. This diffeomorphic parameterization is well suited to cardiac motion modelling, as it encourages smooth and regular deformations. It also enables efficient computation of bidirectional inter-frame deformations. Specifically, using the first-order approximation of the Baker--Campbell--Hausdorff formula~\citep{vercauteren2008symmetric}, the deformation between any two arbitrary frames $\mathbf{I}_i$ and $\mathbf{I}_j$ can be approximated as:
\begin{align}
\label{phi_ex1}
    \mathbf{\Phi}_{j\leftarrow i} & = \exp(\mathbf{V}_{j \leftarrow R} - \mathbf{V}_{i \leftarrow R}), \\
\label{phi_ex2}
    \mathbf{\Phi}_{i\leftarrow j} & = \exp(\mathbf{V}_{i \leftarrow R} - \mathbf{V}_{j \leftarrow R}).
\end{align}
Here, $\mathbf{\Phi}_{j\leftarrow i}$ maps coordinates from $\mathbf{I}_i$ to $\mathbf{I}_j$ and is used to sample $\mathbf{I}_j$ to reconstruct $\mathbf{I}_i$, i.e.,
$\hat{\mathbf{I}}_j(x)=\mathbf{I}_j(\mathbf{\Phi}_{j\leftarrow i}(x))\approx \mathbf{I}_i(x)$.
Similarly, $\mathbf{\Phi}_{i\leftarrow j}$ maps coordinates from $\mathbf{I}_j$ to $\mathbf{I}_i$ and is used to sample $\mathbf{I}_i$ to reconstruct $\mathbf{I}_j$.

\subsection{Latent Trajectory Learning for Cardiac Deformation Representation}
\label{method_rep}
As stated above, given an echocardiography video $\mathbf{I} \in \mathbb{R}^{T \times H \times W}$, we train an autoencoder that predicts velocity fields $\mathbf{V} \in \mathbb{R}^{T \times 2 \times H \times W}$ between each frame and the abstract reference frame. 
The encoder $\mathcal{E}: \mathbb{R}^{H \times W} \rightarrow \mathbb{R} ^D$ maps each  frame to a latent vector $\mathbf{h}_t = \mathcal{E} (\mathbf{I}_t)$ of dimension $D$, while the decoder $\mathcal{D}: \mathbb{R} ^D\rightarrow \mathbb{R}^{2\times H \times W}$ reconstructs the frame-wise velocity field from a latent input $\mathbf{\hat{h}}_t$. As shown in Figure \ref{fig_svf_method}(b), the latent input to the decoder $\mathbf{\hat{h}}_t$ differs from the encoder output $\mathbf{h}_t$ and it's designed to be
\begin{equation}
    \mathbf{\hat{h}}_t = \mathbf{z}^s + \mathbf{z}_t^m, 
\end{equation}
 where $\mathbf{z}^s$ denotes the static component, shared across all frames of the input video, i.e. serving as an anchor in the latent space. $\mathbf{z}_t^m$ represents the motion component, capturing frame-specific deviations around this anchor. 
We constrain $\mathbf{z}_t^m$ to a $M$-dimensional subspace of the latent space spanned by learnable orthogonal basis vectors 
\begin{equation}
    \mathbf{E} = [e_1, e_2, ...e_M] \in \mathbb{R}^{M \times D}, M << D, 
\end{equation} which are optimized jointly with the network parameters using Gram-Schmidt orthonormalization~\citep{wang2024lia}. 
Two 2-layer multilayer perceptrons (MLPs) $\mathbf{f}_1 : \mathbb{R}^D \rightarrow \mathbb{R}^D$ and $\mathbf{f}_2 : \mathbb{R}^D \rightarrow \mathbb{R}^M$, are used to obtain the latent components respectively from encoder output $\mathbf{h}_t$:
\begin{align}
    \mathbf{z}^s & = \mathbf{f}_1 (\frac{1}{T}\sum_{t=1}^T \mathbf{h}_t)\\
    \mathbf{z}_t^m &= \mathbf{E}^\top \mathbf{\alpha}_t
\end{align}
Specifically, we denote the output of $\mathbf{f}_2$ as: 
\begin{equation}
    \alpha_t = \mathbf{f}_2 (\mathbf{h}_t), 
\end{equation}
representing the coordinates of each motion component in the low-dimensional subspace.     
We obtain the frame-wise velocity field from the decoder: 
\begin{equation}
    \mathbf{V}_{t\leftarrow R} = \mathcal{D} ( \mathbf{\hat{h}}_t ) = \mathcal{D} ( \mathbf{z}^s  + \mathbf{z}_t^m),
\end{equation}
and the corresponding deformation field $\Phi_{t \leftarrow R} = \exp(\mathbf{V}_{t \leftarrow R})$. This output reformulation distinguishes ORBIT from LMP~\citep{yang2025latent}. While LMP uses the same latent decomposition to reconstruct image appearance, ORBIT uses it to predict deformation fields via self-supervised registration. Thus, the low-dimensional coordinates $\alpha_t$ are optimized to capture cardiac deformation dynamics, rather than appearance changes.
To train this deformation-based representation, we define a video-level registration loss. Applying Equations~\ref{phi_ex1} and~\ref{phi_ex2}, each frame is registered to up to five subsequent frames in both temporal directions. We use normalized cross-correlation (NCC) as the image similarity term between the fixed frame and the warped moving frame~\citep{luo2010fast}:
\begin{equation}
\label{loss_function}
\begin{aligned}
    \mathcal{L} = - \sum_{f=1}^{5} \sum_{t=1}^{T-f} (& NCC (\mathbf{I}_t \circ \Phi_{t \leftarrow t+f}, \mathbf{I}_{t+f}) \\+ &NCC (\mathbf{I}_{t+f} \circ \Phi_{t+f \leftarrow t}, \mathbf{I}_{t}))
\end{aligned}
\end{equation}
This encourages temporal consistency and smooth deformation dynamics across cardiac cycles.

\subsection{Orientation-Robust Latent Trajectory–Based Cardiac Phase Detection}
\label{method_traj}
After training, only the encoder and the MLP  $\mathbf{f}_2$ are required to analyse cardiac phase dynamics. Each video is represented by a sequence of latent coordinates $\{\alpha_t, t=1,...T\}$,  which form a low-dimensional trajectory that characterizes periodic cardiac motion over time. We investigate very low-dimensional latent motion spaces ($M=1$ and $M=2$), motivated by the low-dimensional and cyclic nature of cardiac motion~\citep{rohe2018low}. The resulting compact trajectories can be directly visualised and analysed to identify cardiac phase transitions without manual phase annotations.
Figure \ref{fig_svf_method}(d) illustrates representative examples of 1D and 2D latent trajectories, where the turning points correspond to ED and ES. The specific association of ED and ES with individual turning points is determined based on the observed relaxation and contraction patterns in the corresponding videos on the validation set.
We detect turning points by locating peaks and valleys in the latent trajectory using the $\texttt{find}\_\texttt{peaks}$ function from $\texttt{scipy}$~\citep{2020SciPy-NMeth}. For model with $M=2$, we select one direction to form a 1D trajectory and perform peak and valley detection.

\section{Experiments and settings}
\label{exp}
\subsection{Datasets}
In this study, we curated a dataset of 4CV view B-mode echocardiography videos acquired during the second trimester of gestation at the John Radcliffe Hospital, Oxford University Hospitals NHS Foundation Trust, UK. All videos were obtained using GE Voluson E10 and E8 ultrasound systems, with a frame resolution of $1136 \times 786$ pixels. The dataset forms part of a retrospective cohort collected under the CAIFE study, which aims to develop AI models for the early detection of congenital heart disease from ultrasound videos~\citep{patey2025prenatal}.

\begin{figure}[]
    \centering
    \includegraphics[width=\columnwidth]{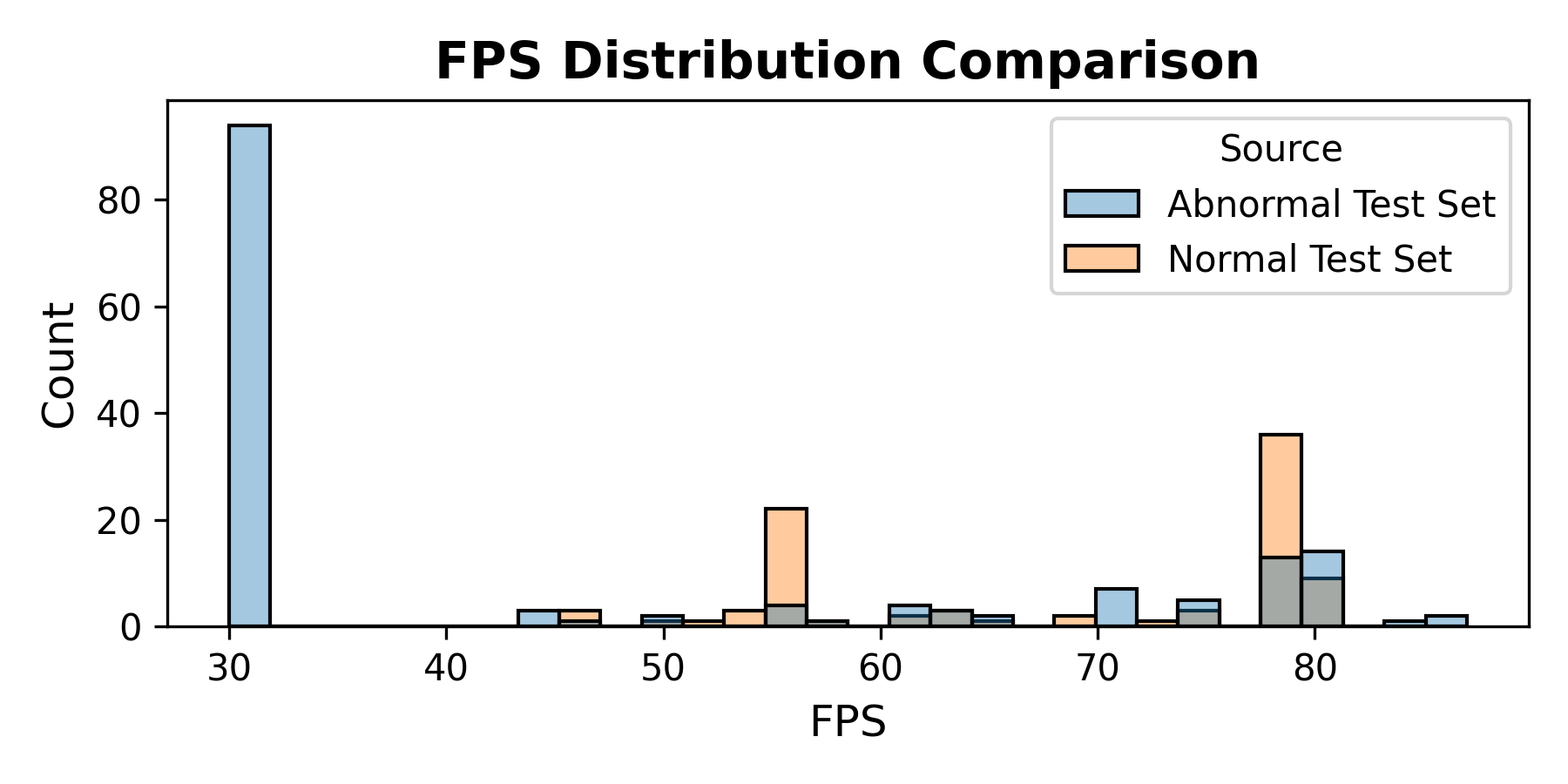}
    \caption{\textbf{Histogram of frames-per-second (FPS) distributions for normal and abnormal test videos.} Blue and orange bars correspond to abnormal and normal cases, respectively; overlapping bins are shown in gray.}
    \label{fig_fps}
\end{figure}

\begin{itemize}
    \item \textbf{Training and validation sets:}
    The training and validation sets comprise 422 videos from 273 healthy participants, each contributing between 1 and 4 videos. The data were split into 85\% for training and 15\% for validation, setting the same configuration for all models presented in this study. No ED or ES annotations are available for either the training or validation sets.
    
    \item \textbf{Normal test set:}  
    The normal test set consists of 88 videos from 78 non-overlapping healthy participants, with 1–2 videos per participant. Frame rates range from 47 to 81 Hz, with a mean of 68 Hz. One experienced fetal cardiologist and one obstetrician with expertise in fetal cardiology independently annotated the ED and ES frames, resulting in a total of 317 annotated frames.  
    
    \item \textbf{Abnormal test set:}  
    The abnormal test set includes 156 videos from 47 participants diagnosed with CHD, covering 13 distinct CHD conditions. Each participant contributed between 1 and 8 videos. Frame rates range from 30 to 87 Hz, with a mean of 48 Hz. The same operators independently annotated the ED and ES frames for these CHD cases, yielding a total of 531 annotated frames. A detailed breakdown of video counts by CHD condition is provided in Table~\ref{tab:chd_num}, and the distribution of frame rates for both normal and abnormal test sets is illustrated in Figure~\ref{fig_fps}.

\end{itemize}

\begin{table}[]
    \centering
    \footnotesize
    \caption{\textbf{Video statistics.} Detailed explanation of CHD abbreviation are presented in \ref{app1}.}
    \begin{tabular}{l l c c}
    \toprule
         Split & Diagnosis & \#Participant & \#Videos  \\
    \midrule
    Train/Validation & Normal & 273 & 422 \\
    \midrule
    Test & Normal & 78 & 88 \\
    \midrule
    \multirow{14}{*}{Test} 
         & CHD & 47 & 156 \\
         \cmidrule{2-4}
         & VSD & 11 & 39 \\
         & CAVSD/B & 10 & 41 \\
         & HLHS & 4 & 17 \\
         & COA & 4 & 7 \\
         & MA/VSD & 4 & 6 \\
         & CAVSD/U & 3 & 13 \\
         & TAVSD/NRA & 3 & 10 \\
         & TAPVC/IC & 2 & 9 \\
         & PA/IVS & 2 & 7\\
         & DTV & 1 & 4 \\
         & FA/T & 1 & 1\\
         & ILA & 1 & 1 \\
         & UVH/RAI & 1 & 1 \\ 
    \bottomrule
    \end{tabular}
    \label{tab:chd_num}
\end{table}

\begin{figure}[]
    \centering
    \includegraphics[width=\columnwidth]{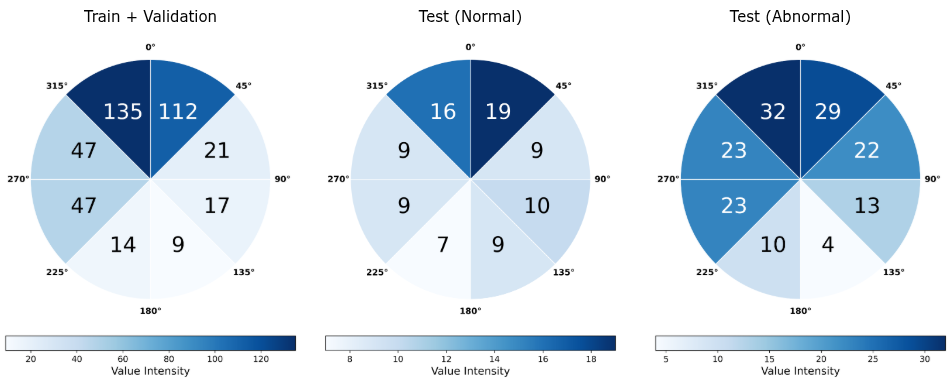}
    \caption{\textbf{Orientation distribution} in train/validation set (normal), test (normal) set and test (abnormal) set. }
    \label{fig_ori_dist_all}
\end{figure}

\paragraph{Orientation distribution:}
An experienced engineer manually annotated the orientation of each video by assigning it to one of eight predefined angle groups, as illustrated in Figure~\ref{fig_ori_count_trainvalid}.
Figure~\ref{fig_ori_dist_all} shows the distribution of heart orientations at the video level for the training/validation, normal test, and abnormal test sets, respectively.

\subsection{Implementation}
For the training and validation sets, the heart region was manually cropped and resized to $156\times156$ pixels, with a temporal downsampling factor of 2 for data preparation. During training, 25 consecutive frames were randomly selected, and a consistent $128\times128$ region was randomly cropped from each video clip as model input. To augment the data, each video was additionally rotated four times clockwise by 0° (original), 90°, 180°, and 270°.
The model was trained using the Adam optimizer with a learning rate of $1e-4$
, a batch size of 16, and for 500 epochs. The localized NCC loss was set with a window size of $9 \times 9$. We used the same latent decomposition architecture as the baseline LMP model~\citep{yang2025latent}\footnote{Model architecture details can be found in the project repository \url{https://github.com/YingyuYyy/CardiacPhase}.}, with an output reformulation to support self-supervised registration, as described in Section~\ref{method}. The model with the lowest validation loss during training was selected for subsequent evaluation. For evaluation on the test sets, the heart region was manually cropped and each video was resized to $128\times128$ pixels without temporal downsampling.

We measured runtime per video by processing all frames jointly in a single forward pass. Across 88 normal test videos ($117.9 \pm 34.9$ frames), latent trajectory prediction required $17.80 \pm 5.73$ ms on an NVIDIA RTX 5000 GPU. Transferring results to the CPU (Intel Xeon w5-2565X) and extracting ED/ES indices added $0.43 \pm 0.16$ ms, for a total of $18.23 \pm 5.74$ ms per video. Measurements excluded data loading and preprocessing and used PyTorch 2.6.0 with CUDA 12.4.

\subsection{Evaluation metrics}
We follow previous work on cardiac phase detection evaluation using the mean absolute error (MAE)~\citep{dezaki2018cardiac,pu2024hfsccd,lu2025optical} between each predicted ED or ES frame and the corresponding ground-truth frames annotated by expert cardiologists. Since multiple ED/ES events occur in each video, the error for each ground-truth frame is computed with respect to its nearest predicted frame. To account for variability in frame rate (frames per second, fps) and heart rate (beats per minute, bpm) across the test videos, we additionally report MAE in milliseconds (MAE-ms) and MAE as a percentage of the cardiac cycle (MAE-\%cycle). In addition, we quantify the detection success rate by applying a 50\% threshold to the MAE-\%cycle metric, representing the proportion of confidently matched ED/ES events. 

\subsection{Statistical analysis}
As cardiac cycles within the same video are highly correlated, the assumption of independence does not hold for event-level analyses (ED or ES). Moreover, each participant may contribute multiple videos acquired under similar anatomical or imaging conditions, making video-level or cycle-level paired testing inappropriate. To address these dependencies, participant-level aggregation is used for non-parametric group comparisons, treating each participant as an independent observational unit. We employ linear mixed-effects models~\citep{lindstrom1988newton} to evaluate statistical significance while leveraging all available data. In these models, the error metric is specified as the dependent variable, method or group are modelled as fixed effects, and random intercepts are included for participant and video to account for intra-participant and intra-video variability. 

\subsection{Baseline methods}
We compare our approach with the original LMP~\citep{yang2025latent} ($M=2$), which applies the latent decomposition for image reconstruction task and requires manual orientation correct towards the apical direction (close to $0\degree$ as illustrated in Figure \ref{fig_ori_count_trainvalid}) before being fed into the deep learning model. In addition, to disentangle the effect of registration-based self-supervision from the effect of rotation augmentation, we retrained the LMP model using the original reconstruction-based self-supervision but with the same rotation augmentation strategy as ORBIT. Specifically, the training clips were randomly rotated by 0$^\circ$, 90$^\circ$, 180$^\circ$, or 270$^\circ$, and no manual orientation correction was applied. We refer to this baseline as \textit{LMP + rotation augmentation} (LMP+rot-aug). This setting provides an augmentation-matched comparison between reconstruction-based and registration-based self-supervision under unknown heart orientations. Table \ref{tab_method_breakdown} summarizes the difference between evaluated methods in this paper. 

\begin{table*}[]
    \centering
    \footnotesize
    \caption{Detailed component breakdown of evaluated methods in our experiments. }
    \label{tab_method_breakdown}
    \begin{tabular}{c c c c c c}
    \toprule
         Method & Orientation Correction & Rotation Augmentation & Latent Decomposition & Reconstruction & Registration\\
         \midrule
         LMP & \checkmark & -- & \checkmark & \checkmark & --\\
         LMP+rot-aug & -- & \checkmark & \checkmark & \checkmark & -- \\
         ORBIT & -- & \checkmark & \checkmark & -- & \checkmark\\
         \bottomrule
    \end{tabular}
\end{table*}

\begin{table*}[t]
  \centering
  \footnotesize
  \caption{\textbf{Cardiac phase detection in fetal echocardiography (88 normal videos).} All models trained with latent motion dimension $M=2$.} Statistical test performed between ORBIT and LMP.  ** $p<0.01$, *** $p<0.001$. 
  \begin{tabular}{c c c c c c c c c c}
    \toprule 
    \multirow{2}{*}{\textbf{Method}} & 
    \multirow{2}{*}{\textbf{Annotator}} & 
    \multicolumn{2}{c}{\makecell{\textbf{MAE (frames)}\\ mean (\textit{std})}} & 
    \multicolumn{2}{c}{\makecell{\textbf{MAE (ms)}\\ mean (\textit{std})}} & 
    \multicolumn{2}{c}{\makecell{\textbf{MAE (\% cycle)}\\ mean (\textit{std})}} & 
    \multicolumn{2}{c}{\makecell{\textbf{\% Samples}\\ MAE/cycle $<$ 50}} 
    \\
    \cmidrule(lr){3-4} \cmidrule(lr){5-6} \cmidrule(lr){7-8} \cmidrule(lr){9-10}
    & & \textbf{ED} & \textbf{ES} & \textbf{ED} & \textbf{ES} & \textbf{ED} & \textbf{ES} & \textbf{ED} & \textbf{ES}\\
    \midrule
    Inter-Observer & \#1 vs \#2 & {1.6 (1.4)} & {1.0 (1.1)} & {23.8 (20.4)} & {14.9 (15.5)} & {6.0 (5.2)} & {3.8 (4.0)} & 100.0 & 100.0\\
    \midrule
    \multirow{2}{*}{\makecell{LMP\\ \citep{yang2025latent}}} 
    & \#1 & {3.4 (6.7)} & {3.7 (7.3)} & {51.1 (102.3)} & {56.4 (107.2)} & {12.9 (26.1)} & {14.3 (27.4)} & 93.4 & 93.1\\
    &  \#2  & {3.8 (6.6)} & {3.4 (7.4)} & {58.5 (100.6)} & {52.1 (109.0)} & {14.7 (25.7)} & {13.1 (27.2)} & 92.4 & 92.7\\
    \midrule
    \multirow{2}{*}{ LMP+rot-aug}
    &  \#1 &  {6.2 (9.4)} &  {6.4 (9.9)} &  {92.1 (136.7)} &  {94.9 (138.7)} &  {23.4 (35.3)}  &  {24.0 (34.5)} &  87.4 &  88.0\\
    &  \#2  &  {6.2 (9.0)} &  {6.2 (9.8)} &  {93.1 (130.9)} &  {92.1 (137.5)} &  {23.4 (32.6)} &  {23.0 (33.7)} &  87.4 &  88.3\\
    \midrule
    \multirow{2}{*}{ORBIT (ours)} 
    & \#1  & {1.9 (3.6)}\sym{**} & {1.6 (2.6)}\sym{***} & {28.2 (53.0)}\sym{**} & {23.5 (41.7)}\sym{***} & {7.2 (13.8)}\sym{**} & {6.0 (10.8)}\sym{***} & 98.0 & 98.7\\
    & \#2  & {2.4 (3.4)}\sym{**} & {1.5 (2.5)}\sym{***} & {35.7 (49.0)}\sym{**} & {22.5 (39.9)}\sym{***} & {9.0 (12.4)}\sym{**} & {5.6 (10.2)}\sym{***} & 97.8 & 98.7\\
    \bottomrule
  \end{tabular}
  
  \label{tab_normal_result}
\end{table*}

\subsection{Experiments}
We perform the following experiments to demonstrate the effectiveness and robustness of the proposed ORBIT method. 

\subsubsection{In-distribution experiments}
We first evaluated the performance of all evaluated methods on the normal test data. In addition, we compared LMP+rot-aug and our ORBIT against different heart orientation groups to study the orientation robustness.

\subsubsection{Out-of-distribution experiments}
Since the model  was trained entirely on normal fetal echocardiography videos, we investigated its robustness under distributional shift, in particular, its performance on cases with CHD conditions. Additionally, we tested the influence of an automatic heart cropping tool \citep{yang2025deep} for video preprocessing. As the region of interest (ROI) was automatically detected using a deep learning-based cropping tool, which can occasionally mislocalize the heart (e.g., incomplete, oversized, or undersized crops), we further evaluated robustness by comparing performance between automatically and manually cropped frames.

\section{Results}
\label{result}

\begin{figure}[]
    \centering
    \includegraphics[width=\columnwidth]{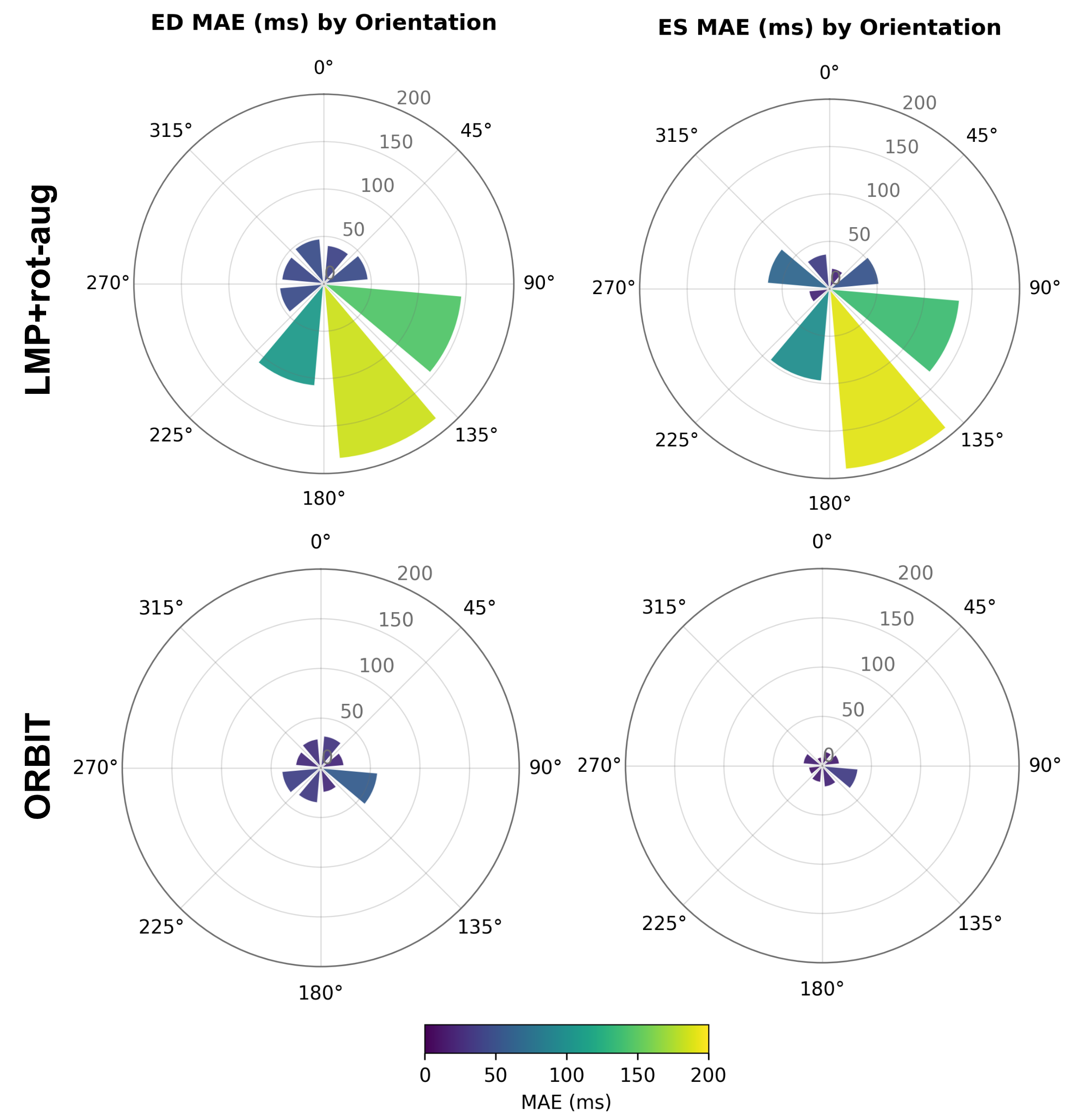}
    \caption{\textbf{Cardiac phase detection MAE (ms) grouped by heart orientation} for LMP+rot-aug and ORBIT. Polar bar plots show the mean MAE within each orientation group, with separate plots for ED and ES detection. The quantitative summary is provided in Table~\ref{tab_angle_err_d1}.}
    \label{fig_d1_angle_error}
\end{figure}

\begin{figure*}[]
    \centering
    \includegraphics[width=\textwidth]{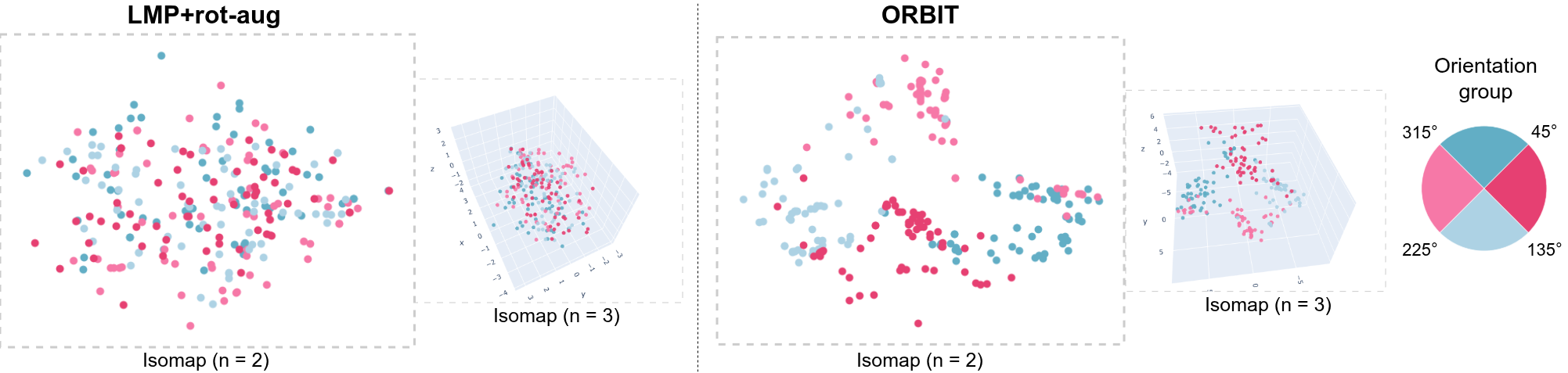}
    \caption{\textbf{Isomap dimension reduction visualization of the static component $\mathbf{z}^s$ on validation dataset (63 videos * 4 rotations).}}
    \label{fig_d1_isomap}
\end{figure*}

\begin{figure}[]
    \centering
    \includegraphics[width=\columnwidth]{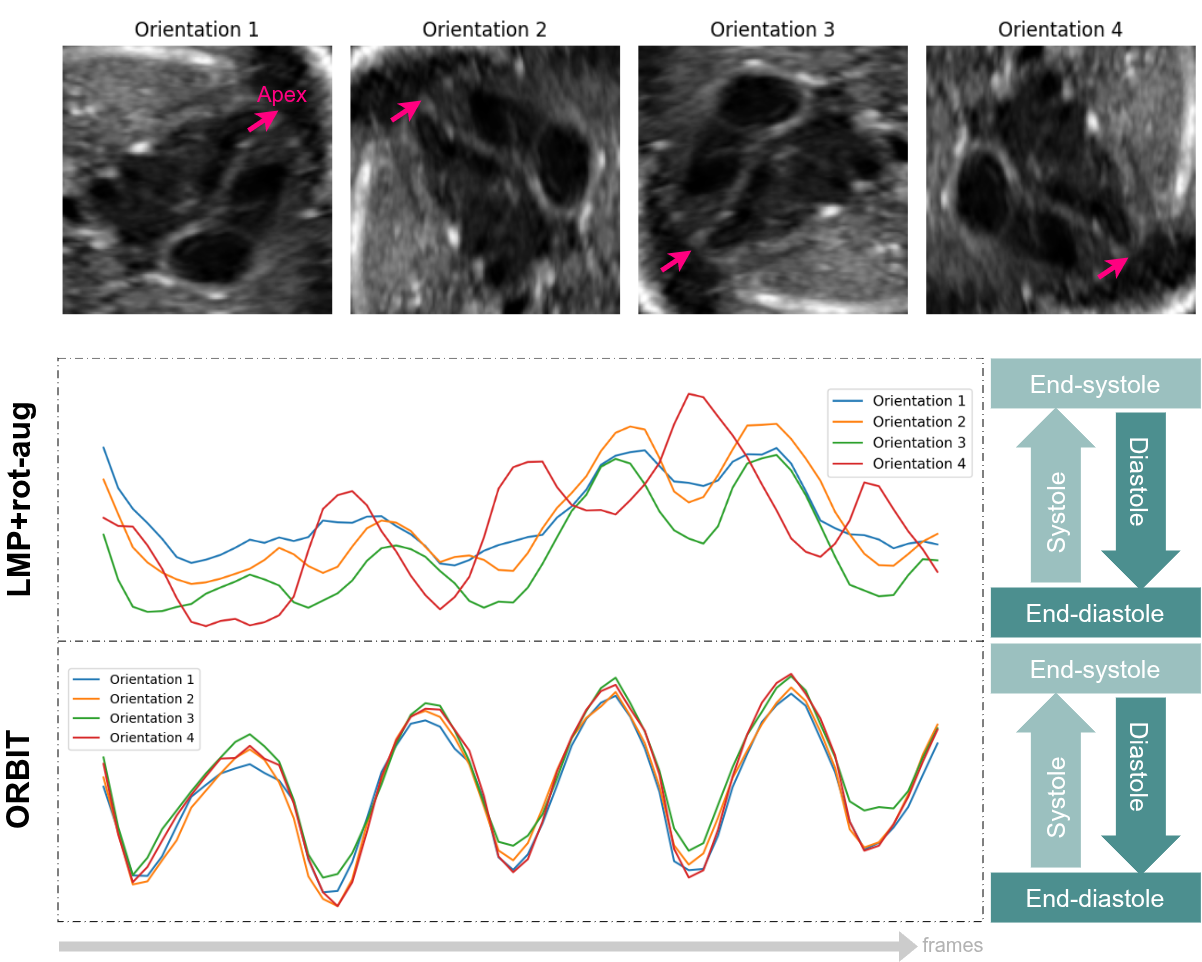}
    \caption{\textbf{Latent motion curve (1D) consistency across orientations.} Four orientations of the same video are shown in the first row, with red arrows indicating the apex location. The second row shows 1D latent trajectories from the LMP+rot-aug encoder, where one trajectory (red) exhibits an inverted trend. The third row shows ORBIT trajectories, which are consistent across orientations. Corresponding video illustrations of the LMP+rot-aug and ORBIT latent trajectories are provided in the supplementary material.}
    \label{fig_d1_example}
\end{figure}

\subsection{In-distribution results}
\subsubsection{Performance on Normal Test Data} Table~\ref{tab_normal_result} presents the evaluation results on the normal test videos. The proposed ORBIT framework consistently and significantly outperforms the baseline method LMP and its variant LMP+rot-aug in detecting both ED and ES events, measured by MAE in frames, milliseconds, and percentage of the cardiac cycle. 
In particular, ORBIT reduces the mean error in cycle percentage by 5\% for ED and 7\% for ES compared with LMP, achieving prediction accuracy closer to inter-observer variability. Moreover, ORBIT does not rely on any orientation information for cardiac phase prediction, making it substantially more versatile than the baseline LMP model, which requires manual rotation of input videos towards the apical orientation. Although LMP+rot-aug uses the same rotation augmentation, its MAE is three times higher than that of ORBIT. We provide further analysis in the following section.

\subsubsection{Orientation Robustness}
The key improvement of ORBIT lies in its orientation robustness. In the context of latent-trajectory-based unsupervised phase detection, we define orientation robustness as maintaining a consistent association between cardiac phases and trajectory turning points across different heart orientations. Specifically, the same type of turning point, such as a peak or valley, should correspond to ED or ES regardless of the heart orientation of the input video. ORBIT therefore enables cardiac phases to be inferred using a fixed trajectory interpretation, without requiring orientation-dependent reassignment of peaks and valleys.

For the orientation robustness analysis, we use the LMP+rot-aug baseline rather than the original LMP setting. This baseline retains the reconstruction-based self-supervision of LMP, but is retrained with the same rotation augmentation strategy as ORBIT without manual orientation correction. For simplicity, interpretability, and direct visualization, we choose $M=1$ as the latent dimension to compare the performance between LMP+rot-aug and ORBIT in Figures~5--7. A detailed analysis of the choice of latent dimensionality is provided in Section \ref{sec_m12}.

As shown in Figure \ref{fig_d1_angle_error} (quantitative summary in Table \ref{tab_angle_err_d1}), the latent trajectory learned by LMP+rot-aug exhibits strong orientation dependency. Errors between 90° and 225° increase remarkably for both ED and ES. In contrast, ORBIT maintains a nearly flat error distribution across all orientations, demonstrating robustness.

We illustrate an example in Figure \ref{fig_d1_example}: when visualizing the 1D latent trajectories for one validation case under four orientations, LMP+rot-aug produces an inverted trajectory at one orientation (in red), effectively swapping the meanings of ED and ES. This inversion indicates that reconstruction-based models require explicit orientation information to interpret phase correctly. One unified assignment between trajectory peak/valley and ED/ES is not possible. ORBIT, however, yields consistent trajectories across all orientations, reflecting true orientation invariance due to its modelling of cardiac deformation.

To investigate the reason for this robustness, we visualize the static component $\mathbf{z}^s$ using Isomap dimensionality reduction on the validation set (Figure~\ref{fig_d1_isomap}). The reconstruction-based model(LMP+rot-aug) shows mixed embeddings across orientation groups, while the registration-based ORBIT produces manifolds where static components are naturally grouped by orientation. This disentanglement removes the influence of heart orientation, allowing the latent motion component to capture a stable and consistent motion pattern that is explainable at group level, resulting in orientation-robust phase detection.

\begin{figure*}[h]
    \centering
    \includegraphics[width=\textwidth]{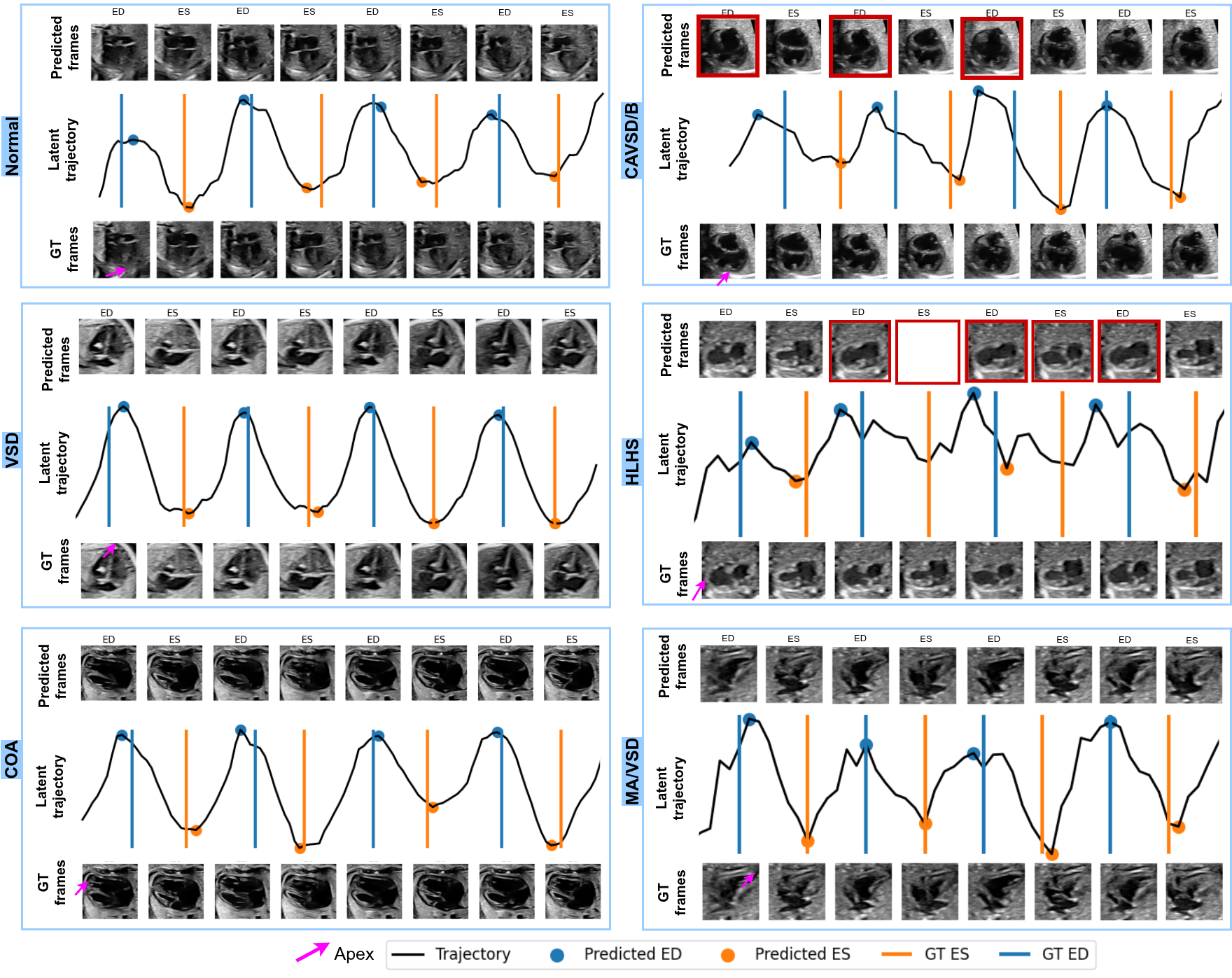}
    \caption{\textbf{Prediction examples using ORBIT.} Red box indicates MAE larger than 2 frames.}
    \label{fig_example_all}
\end{figure*}

\begin{table*}[t]
  \centering
  \footnotesize
  \caption{\textbf{Cardiac phase detection in fetal echocardiography (156 CHD videos).} All models trained with latent motion dimension $M=2$.Statistical test performed between ORBIT and LMP.  * $p<0.05$, ** $p<0.01$, *** $p<0.001$. }
  \begin{tabular}{c c c c c c c c c c}
    \toprule
    \multirow{2}{*}{\textbf{Self-supervision}} & 
    \multirow{2}{*}{\textbf{Annotator}} & 
    \multicolumn{2}{c}{\makecell{\textbf{MAE (frames)}\\ mean (\textit{std})}} & 
    \multicolumn{2}{c}{\makecell{\textbf{MAE (ms)}\\ mean (\textit{std})}} & 
    \multicolumn{2}{c}{\makecell{\textbf{MAE (\% cycle)}\\ mean (\textit{std})}} & 
    \multicolumn{2}{c}{\makecell{\textbf{\% Samples}\\ MAE/cycle $<$ 50}} 
    \\
    \cmidrule(lr){3-4} \cmidrule(lr){5-6} \cmidrule(lr){7-8} \cmidrule(lr){9-10}
    & & \textbf{ED} & \textbf{ES} & \textbf{ED} & \textbf{ES} & \textbf{ED} & \textbf{ES} & \textbf{ED} & \textbf{ES}\\
    \midrule
    Inter-Observer & \#1 vs \#2 & {0.8 (1.0)} & {0.8 (1.0)} & {16.9 (20.8)} & {19.2 (22.5)} & {4.2 (5.1)} & {4.8 (6.0)} & 100.0 & 100.0\\
    \midrule
    \multirow{2}{*}{\makecell{LMP\\ \citep{yang2025latent}}}
    & \#1 & {3.3 (6.8)} & {3.3 (7.1)} & {74.2 (133.8)} & {74.3 (140.0)} & {18.4 (33.4)} & {18.0 (32.9)} & 90.2 & 89.8\\
    & \#2  & {3.5 (6.7)} & {3.2 (7.1)} & {78.6 (132.0)} & {70.6 (138.6)} & {19.8 (34.1)} & {17.4 (33.5)} & 90.8 & 89.0\\
    \midrule
    \multirow{2}{*}{ LMP+rot-aug}
    &  \#1 &  {5.3 (7.5)} &  {5.2 (7.1)} &  {118.9 (148.4)} &  {118.8 (154.0)} &  {29.5 (37.7)} &  {29.1 (37.4)} &  84.7 &  82.3\\
    &  \#2  &  {5.5 (7.7)} &  {5.2 (7.2)} &  {121.6 (148.5)} &  {118.4 (153.7)} &  {30.7 (38.6)} &  {29.6 (38.2)} &  82.5 &  81.0\\
    \midrule
    \multirow{2}{*}{ORBIT (ours)} 
    & \#1  & {2.4 (3.6)}\sym{**} & {2.1 (3.8)}\sym{***} & {56.0 (75.0)} & {47.3 (81.6)}\sym{*}  & {13.9 (19.0)}\sym{*} & {11.8 (20.7)}\sym{**}  & 95.3 & 95.3\\
    & \#2  & {2.5 (3.6)}\sym{**} & {2.0 (3.8)}\sym{***}  & {59.5 (77.5)} & {48.1 (82.3)}\sym{*}  & {14.9 (19.8)}\sym{*} & {12.2 (21.4)}\sym{**}  & 94.7 & 95.3\\
    \bottomrule
  \end{tabular}
  
\label{tab_chd_result}
\end{table*}

\subsection{Out-of-distribution results}
\subsubsection{Performance on Abnormal Test Data}
Table \ref{tab_chd_result} summarizes the overall cardiac phase detection errors on abnormal (CHD) test data for both the baseline LMP model, LMP+rot-aug and the proposed ORBIT model. Consistent with the normal-case results, ORBIT achieves lower MAE in both frames and cardiac-cycle percentage compared with LMP and LMP+rot-aug.

To further assess robustness across physiological conditions, we compared the participant-level MAE (median/mean of error in frames) between normal and abnormal groups for each method and each phase using the Mann–Whitney U test. The differences between groups were not statistically significant ($p > 0.05$ for both median and mean) for either LMP or ORBIT, indicating that both latent-trajectory models maintain comparable accuracy on abnormal fetal hearts.

\begin{table}[t]
  \centering
  \footnotesize
  \caption{\textbf{Evaluation between normal and different CHD conditions using ORBIT method. (\#2 Annotator, latent motion dimension $M=2$)}. Statistical comparison performed between Normal group and VSD, CAVSD/B group respectively. * $p<0.05$.
  }
  \begin{tabular}{c c c c c}
    \toprule 
    \multirow{2}{*}{\textbf{Group}} & 
    \multicolumn{2}{c}{\makecell{\textbf{MAE (frames)}\\ mean (\textit{std})}} &
    \multicolumn{2}{c}{\makecell{\textbf{MAE (\% cycles)}\\ mean (\textit{std})}} 
    \\
    \cmidrule(lr){2-3}  \cmidrule(lr){4-5}
    & \textbf{ED} & \textbf{ES} & \textbf{ED} & \textbf{ES}\\
    \midrule
    Normal & {2.4 (3.4)} & {1.5 (2.5)} & {9.0 (12.4)} & {5.6 (10.2)}\\
    \midrule
    CHD &  {2.5 (3.6)} & {2.0 (3.8)} & {14.9 (19.8)} & {12.2 (21.4)} \\
    \midrule 
    VSD & {1.6 (2.5)} & {1.6 (3.4)} & {9.1 (12.4)} & {9.0 (16.4)}\\
    CAVSD/B & {3.0 (3.7)}\sym{*} & {1.8 (3.1)} & {16.9 (21.7)}\sym{*} & {9.9 (17.1)}\\
    \midrule
    HLHS & {3.9 (6.7)} & {3.0 (5.7)} & {20.9 (26.8)} & {20.5 (33.6)}  \\
    COA & {2.1 (1.0)} & {2.0 (3.9)} & {7.8 (3.3)} & {7.5 (12.6)}\\
    MA/VSD & {2.0 (2.0)} & {1.2 (1.2)} & {14.7 (16.6)} & {7.5 (8.8)}\\
    \bottomrule
  \end{tabular}
  \label{tab_chd_detail}
\end{table}

Although the overall degradation in MAE for abnormal cases is not statistically significant at the group level, we are interested whether ORBIT’s performance varies across different CHD subtypes. Table~\ref{tab_chd_detail} reports the subtype-level MAE for CHD conditions with available test cases. Given the limited number of participants in several diagnostic groups, this analysis is intended to be exploratory.

Among the larger subgroups, ORBIT shows stable performance in VSD, with MAE values comparable to those observed in the normal cohort. In contrast, larger deviations are observed in CAVSD/B, particularly for ED prediction (16.9\% vs. 9.0\% of the cardiac cycle in the normal cohort). For smaller subgroups such as HLHS, COA, and MA/VSD, their descriptive results suggest that performance may vary across CHD morphologies, with larger errors observed in some complex conditions such as HLHS.
The observed variation may reflect differences in underlying anatomical and motion characteristics. VSD and mild forms of COA may preserve biventricular structure and global contraction patterns that are closer to those seen in normal hearts, which could allow ORBIT to generalize from normal training data. In contrast, CAVSD/B and HLHS can involve substantially altered ventricular morphology and atypical motion dynamics, which may reduce the salience of the deformation cues used by ORBIT to extract latent motion trajectories. As a result, ED and ES turning points may become less well defined in the latent space, leading to increased prediction error. Representative latent trajectory examples are shown in Fig.~\ref{fig_example_all}.

Together, these results suggest that ORBIT maintains stable performance across the overall CHD cohort, but the subtype-level analysis remains underpowered. Except for subgroups with sufficient participant numbers, the reported subtype results should be interpreted as descriptive observations rather than statistical evidence of disease-specific performance differences. Larger and more balanced CHD cohorts will be required to determine how ORBIT performs across individual CHD morphologies and whether complex ventricular morphology or dysfunction systematically affects ED/ES detection.

\subsubsection{Influence of Frame Rate}
As shown in Table \ref{tab_chd_result}, a modest degradation is observed when errors are expressed in milliseconds or cardiac-cycle percentage on the abnormal test set. This likely reflects acquisition characteristics: the abnormal dataset is captured at a lower frame rate (Figure~\ref{fig_fps}), with many cases recorded at 30 fps. Because time-based metrics scale with frame interval, lower fps leads to larger apparent temporal errors. This discrepancy arises from retrospective data storage practices, where high-frame-rate acquisitions are often down-sampled to 30 fps to reduce file size.

This frame-rate difference also helps explain the lower inter-observer MAE observed for CHD cases (Table \ref{tab_chd_result}) when measured in frames than that in normal cases (Table \ref{tab_normal_result}) The higher frame rates in the normal cohort provide more visually similar candidate frames for the same cardiac phase, allowing greater frame-level disagreement, whereas the lower frame rates in the CHD cohort constrain this variability. Accordingly, the difference is reduced when expressed in milliseconds. In addition, cohort composition may also contribute to this discrepancy, as the normal test set contains more participants and potentially greater variation in heart appearance and image quality. Thus, the lower frame-level inter-observer MAE in CHD cases should be interpreted cautiously and does not necessarily imply that CHD phase annotation is easier.
\begin{table}[t]
  \centering
  \footnotesize
  \caption{\textbf{Evaluation on abnormal test data under different cropping strategies (\#2 Annotator, latent motion dimension $M=2$).} Statistical comparison performed between crop strategies for each method. Stars: * $p<0.05$, ** $p<0.01$, *** $p<0.001$.}
  \begin{tabular}{c c c c}
    \toprule 
    \multirow{2}{*}{\textbf{Method}} & 
    \multirow{2}{*}{\textbf{Crop Strategy}} & 
    \multicolumn{2}{c}{\makecell{\textbf{MAE (frames)}\\ mean (\textit{std})}} 
    \\
    \cmidrule(lr){3-4}
    & & \textbf{ED} & \textbf{ES} \\
    \midrule
    \multirow{2}{*}{LMP} 
    & Manual & {3.5 (6.7)} & {3.2 (7.1)}\\
    & FEHT & {4.4 (6.8)}\sym{***} & {4.0 (7.2)}\sym{***}\\
    \midrule
    \multirow{2}{*}{ LMP+rot-aug} 
    &  Manual &  {5.5 (7.7)} &  {5.2 (7.2)}\\
    &  FEHT &  {7.2 (12.5)}\sym{***} &  {7.1 (12.9)}\sym{***}\\
    \midrule
    \multirow{2}{*}{ORBIT (ours)} 
    & Manual  & {2.5 (3.6)} & {2.0 (3.8)} \\
    & FEHT  & {3.5 (4.9)}\sym{***} & {3.0 (4.9)}\sym{***} \\
    \bottomrule
  \end{tabular}
  \label{tab_crop}
\end{table}

\subsubsection{Effect of Cropping Variations}
To assess robustness to preprocessing differences, we compared performance between manually cropped and automatically (tool-based) cropped images using the FEHT method~\citep{yang2025deep}, as summarized in Table~\ref{tab_crop}.
For both LMP, LMP+rot-aug and ORBIT, automatic cropping increased the mean error by approximately one frame on the abnormal test set. This degradation is primarily attributed to imperfect alignment of the automatically extracted heart region with the manually defined region used during training.
The results highlight that consistent and accurate localization of the cardiac region remains critical for reliable phase inference in latent-trajectory–based methods. Because ORBIT relies on capturing subtle motion patterns within the heart region, the cropped area should have sufficient resolution and minimal background interference.

\begin{table*}[]
 
  \centering
  \footnotesize
  \caption{ORBIT model with latent subspace dimension $M=1,2$ on ED/ES detection (\#2 Annotator.) Statistical test conducted between model of $M=1$ and $M=2$ for each test set. Stars: * $p<0.05$, ** $p<0.01$, *** $p<0.001$.}
  \label{tab_deform12}
   \begin{tabular}{c c c c c c c c c c}
    \toprule
    \multirow{2}{*}{\textbf{Test-set}} & 
    \multirow{2}{*}{\textbf{M}} & 
    \multicolumn{2}{c}{\makecell{\textbf{MAE (frames)}\\ mean (\textit{std})}} & 
    \multicolumn{2}{c}{\makecell{\textbf{MAE (ms)}\\ mean (\textit{std})}} & 
    \multicolumn{2}{c}{\makecell{\textbf{MAE (\% cycle)}\\ mean (\textit{std})}} & 
    \multicolumn{2}{c}{\makecell{\textbf{\% Samples}\\ MAE/cycle $<$ 50}} 
    \\
    \cmidrule(lr){3-4} \cmidrule(lr){5-6} \cmidrule(lr){7-8} \cmidrule(lr){9-10}
    & & \textbf{ED} & \textbf{ES} & \textbf{ED} & \textbf{ES} & \textbf{ED} & \textbf{ES} & \textbf{ED} & \textbf{ES} \\
    \midrule
    \multirow{2}{*}{Normal} 
    & 1  & {2.3 (2.5)} & {1.2 (1.8)} & {33.4 (35.0)} & {17.8 (27.5)} & {8.5 (8.8)} & {4.5 (7.3)} & 99.0 & 99.7\\
    & 2  & {2.4 (3.4)} & {1.5 (2.5)}\sym{*} & {35.7 (49.0)} & {22.5 (39.9)} & {9.0 (12.4)} & {5.6 (10.2)} & 98.0 & 98.7\\
    \midrule
    \multirow{2}{*}{Abnormal} 
    & 1 & {2.6 (2.9)} & {2.3 (4.0)} & {65.9 (78.6)} & {57.1 (91.1)} & {16.4 (19.8)} & {14.5 (23.5)} & 94.0 & 93.0\\
    & 2 & {2.5 (3.6)}\sym{**} & {2.0 (3.9)}\sym{**} & {59.5 (77.5)}\sym{***} & {48.1 (82.3)}\sym{**} & {14.9 (19.7)}\sym{***} & {12.2 (21.4)}\sym{**} & 94.7 & 95.3\\
    \bottomrule
  \end{tabular}
~
 
  \centering
  \footnotesize
  \caption{LMP+rot-aug model with latent subspace dimension $M=1,2$ on ED/ES detection (\#2 Annotator)}
  \label{tab_lmp-rot-aug12}
   \begin{tabular}{c c c c c c c c c c}
    \toprule
    \multirow{2}{*}{\textbf{Test-set}} & 
    \multirow{2}{*}{\textbf{M}} & 
    \multicolumn{2}{c}{\makecell{\textbf{MAE (frames)}\\ mean (\textit{std})}} & 
    \multicolumn{2}{c}{\makecell{\textbf{MAE (ms)}\\ mean (\textit{std})}} & 
    \multicolumn{2}{c}{\makecell{\textbf{MAE (\% cycle)}\\ mean (\textit{std})}} & 
    \multicolumn{2}{c}{\makecell{\textbf{\% Samples}\\ MAE/cycle $<$ 50}} 
    \\
    \cmidrule(lr){3-4} \cmidrule(lr){5-6} \cmidrule(lr){7-8} \cmidrule(lr){9-10}
    & & \textbf{ED} & \textbf{ES} & \textbf{ED} & \textbf{ES} & \textbf{ED} & \textbf{ES} & \textbf{ED} & \textbf{ES} \\
    \midrule
    \multirow{2}{*}{Normal} 
    & 1  & {5.1 (6.9)} & {4.5 (7.3)} & {74.7 (98.2)} & {67.1 (109.7)} & {18.8 (24.7)} & {16.9 (28.1)} & 89.5 & 89.9\\
    & 2  & {6.2 (9.0)} & {6.2 (9.8)} & {93.1 (130.9)} & {92.1 (137.5)} & {23.4 (32.6)} & {23.0 (33.7)} & 87.4 & 88.3\\
    \midrule
    \multirow{2}{*}{Abnormal} 
    & 1 & {4.6 (6.7)} & {4.3 (7.5)} & {104.9 (125.8)} & {92.5 (132.6)} & {26.4 (32.1)} & {22.8 (33.0)} & 84.6 & 84.4\\
    & 2 & {5.5 (7.7)} & {5.2 (7.2)} & {121.6 (148.5)} & {118.4 (153.7)} & {30.7 (38.6)} & {29.6 (38.2)} & 82.5 & 81.0\\
    \bottomrule
  \end{tabular}
~
 
  \centering
  \footnotesize
  \caption{LMP model with latent subspace dimension $M=1,2$ on ED/ES detection (\#2 Annotator)}
  \label{tab_lmp12}
   \begin{tabular}{c c c c c c c c c c}
    \toprule
    \multirow{2}{*}{\textbf{Test-set}} & 
    \multirow{2}{*}{\textbf{M}} & 
    \multicolumn{2}{c}{\makecell{\textbf{MAE (frames)}\\ mean (\textit{std})}} & 
    \multicolumn{2}{c}{\makecell{\textbf{MAE (ms)}\\ mean (\textit{std})}} & 
    \multicolumn{2}{c}{\makecell{\textbf{MAE (\% cycle)}\\ mean (\textit{std})}} & 
    \multicolumn{2}{c}{\makecell{\textbf{\% Samples}\\ MAE/cycle $<$ 50}} 
    \\
    \cmidrule(lr){3-4} \cmidrule(lr){5-6} \cmidrule(lr){7-8} \cmidrule(lr){9-10}
    & & \textbf{ED} & \textbf{ES} & \textbf{ED} & \textbf{ES} & \textbf{ED} & \textbf{ES} & \textbf{ED} & \textbf{ES} \\
    \midrule
    \multirow{2}{*}{Normal} 
    & 1  & {9.1 (10.0)} & {9.0 (9.7)} & {136.6 (149.8)} & {134.5 (140.7)} & {34.3 (37.7)} & {34.0 (35.7)} & 74.1 & 75.7\\
    & 2  & {3.8 (6.6)} & {3.4 (7.4)} & {58.5 (100.6)} & {52.1 (109.0)} & {14.7 (25.7)} & {13.1 (27.2)} & 92.4 & 92.7\\
    \midrule
    \multirow{2}{*}{Abnormal} 
    & 1 & {7.2 (7.3)} & {7.8 (8.1)} & {156.4 (127.1)} & {169.5 (136.4)} & {39.2 (32.2)} & {42.3 (34.3)} & 71.9 & 68.7\\
    & 2 & {3.5 (6.7)} & {3.2 (7.1)} & {78.6 (132.0)} & {70.6 (138.6)} & {19.8 (34.1)} & {17.4 (33.5)} & 90.8 & 89.0\\
    \bottomrule
  \end{tabular}
\end{table*}

\subsection{Influence of Latent Dimension}
\label{sec_m12}
We evaluated each method using latent dimensions of $M=1$ and $M=2$ (LMP: Table~\ref{tab_lmp12}; LMP+rot-aug: Table~\ref{tab_lmp-rot-aug12}; ORBIT: Table~\ref{tab_deform12}). The effect of reducing the latent dimension differed markedly across the three methods. For the original LMP model, $M=2$ substantially outperformed $M=1$ on both the normal and abnormal test sets. We conjecture that, because LMP learns directly from video appearance and is trained on a narrow range of heart orientations, a one-dimensional representation may not fully disentangle cardiac contraction--relaxation from nuisance appearance variations, including fetal motion, probe motion, and ultrasound artefacts. A second latent dimension may provide additional capacity to capture these factors while preserving cardiac phase information.

In contrast, for LMP+rot-aug, $M=1$ consistently outperformed $M=2$. When appearance-based learning is combined with rotation augmentation, the more constrained one-dimensional latent space may encourage cardiac motion observed at different orientations to be represented along a common motion axis. A two-dimensional space provides greater freedom for orientation-dependent or nuisance variation to be encoded, potentially reducing the consistency of the trajectories used to identify ED and ES. The one-dimensional representation also permits direct visualization of trajectory alignment and inversion under image rotation.

ORBIT was substantially less sensitive to the choice of latent dimensionality. On the normal test set, $M=1$ achieved slightly lower errors, including a statistically significant reduction in ES frame MAE, whereas $M=2$ produced significantly lower ED and ES errors on the abnormal test set. The latter may reflect the greater capacity of a two-dimensional deformation subspace to represent more complex or atypical cardiac motion. Nevertheless, the differences between $M=1$ and $M=2$ were considerably smaller in ORBIT than those observed for the original LMP model and its variant LMP+rot-aug. This indicates that learning latent cardiac deformation, rather than global video appearance, reduces contamination by non-cardiac motion and makes the representation more robust to the choice of latent dimension.

\begin{table*}[]
    \centering
    \footnotesize
    \caption{CHD condition names.}
    \begin{tabular}{l l}
    \toprule
    Abbreviation & Full name \\
    \midrule
    CAVSD/B & Complete atrioventricular septal defect with balanced ventricles\\
    CAVSD/U & Complete atrioventricular septal defect with unbalanced ventricles \\
    COA & Coarctation of the aorta\\
    DTV & Dysplastic tricuspid valve\\
    FA/T & Fetal Arrhythmias/Tachycardia\\
    HLHS & Hypoplastic left heart syndrome \\
    ILA & Isolated left isomerism\\
    MA/VSD & Mitral atresia with ventricular septal defect\\
    PA/IVS & Pulmonary atresia with the intact interventricular septum\\
    TAPVC/IC & Total anomalous pulmonary venous connections (infra-cardiac type)\\
    TAVSD/NRA & Tricuspid atresia with ventricular septal defects and normally related great arteries (NRA)\\
    UVH/RAI & Univentricular heart with right atrial isomerism (RAI)\\
    VSD & Moderate/large ventricular septal defects\\
    \bottomrule
    \end{tabular}
    \label{tab:chd_name}

\centering
\footnotesize
\caption{Orientation-wise cardiac phase detection mean absolute error (ms) on normal test sets (\#2 annotator). Group 1: 0\degree-45\degree, Group 2: 45\degree-90\degree, Group 3: 90\degree-135\degree, Group 4: 135\degree-180\degree, Group 5: 180\degree-225\degree, Group 6: 225\degree-270\degree, Group 7: 270\degree-315\degree, Group 8: 315\degree-360\degree. Readers could find the number of cases under each cardiac orientation angle from Figure \ref{fig_ori_dist_all}.}
\label{tab_angle_err_d1}
\begin{tabular}{c c c c c c c c c c c}
    \toprule
    \textbf{Method} & \textbf{Phase} & \textbf{Group 1} & \textbf{Group 2} & \textbf{Group 3} & \textbf{Group 4} & \textbf{Group 5} & \textbf{Group 6} & \textbf{Group 7} & \textbf{Group 8}\\
    \midrule
     \multirow{2}{*}{LMP+rot-aug} & ED & {40.9 \tiny(32.8)} & {47.0 \tiny(37.4)} & {146.0 \tiny(97.8)} & {185.0 \tiny(147.6)} & {108.2 \tiny(140.1)} & {47.0 \tiny(24.6)} & {44.6 \tiny(75.6)} & {47.8 \tiny(88.6)} \\
     &  ES & {22.0 \tiny(19.8)} & {52.3 \tiny(70.2)} & {137.5 \tiny(141.2)} & {191.2 \tiny(153.7)} & {97.8 \tiny(152.4)} & {22.1 \tiny(13.8)} & {65.9 \tiny(105.3)} & {37.1 \tiny(74.5)}\\
     \midrule
     \multirow{2}{*}{ORBIT} & ED & {32.5 \tiny(19.3)} & {23.8 \tiny(20.2)} & {57.8 \tiny(83.9)} & {25.0 \tiny(15.0)} & {35.6 \tiny(17.9)} & {39.6 \tiny(22.1)} & {25.6 \tiny(16.9)} & {29.5 \tiny(26.1)}\\
     & ES & {14.9 \tiny(14.4)} & {17.8 \tiny(16.1)} & {36.6 \tiny(67.2)} & {22.0 \tiny(19.2)} & {17.2 \tiny(14.5)} & {14.2 \tiny(14.1)} & {19.8 \tiny(20.7)} & {9.0 \tiny(11.0)}\\
     \bottomrule
\end{tabular}
\end{table*}

\section{Discussion and Conclusion}
\label{dis}
In this work, we proposed ORBIT (Orientation-Robust Beat Inference from Trajectories), a self-supervised framework for annotation-free cardiac phase detection in fetal four-chamber echocardiography videos. Building on latent trajectory-based phase modelling, ORBIT reformulates the self-supervised learning target from image reconstruction to inter-frame registration. Specifically, the model learns to decompose stationary velocity fields that parameterize frame-wise cardiac deformation. The resulting low-dimensional latent motion component captures cyclical contraction-relaxation dynamics, from which ES and ED frames can be inferred without manual phase annotations.
Trained solely on normal fetal echocardiography videos, ORBIT demonstrates consistent performance on both normal and abnormal (CHD) data without requiring prior heart orientation information. Compared with reconstruction-based latent trajectory methods that rely on explicit orientation alignment or exhibit orientation-dependent trajectory interpretation, ORBIT learns a deformation-based latent trajectory that is more stable across diverse fetal heart orientations. These findings suggest that registration-based latent motion modelling can provide a useful foundation for real-world fetal video analysis, including heartbeat segmentation, video summarization, and other downstream tasks in fetal echocardiography interpretation.
Disease-wise evaluation further revealed that ORBIT maintains stable accuracy across most common CHD types in the cohort, while showing increased sensitivity in certain severe cases. Although limited sample sizes of CHD cases prevent definitive conclusions, these findings suggest that latent motion trajectory modelling may be used for capturing disease-specific motion signatures. Future work will investigate this direction to explore the potential of ORBIT for prenatal congenital heart disease characterization and detection.

\appendix
\section{Congenital heart diseases included in our study}
\label{app1}
We present the full description of each congenital heart disease included in our study in Table \ref{tab:chd_name}.

\section{Quantitative results of LMP+rot-aug}
We present in Table \ref{tab_angle_err_d1} the quantitative MAE for each heart orientation group using MLP+rot-aug and ORBIT method ($M=1$), which complements the results shown in Figure \ref{fig_d1_angle_error}.

\FloatBarrier 
\section*{Acknowledgement}
This work was supported by the InnoHK-funded Hong Kong Centre for Cerebro-cardiovascular Health Engineering (COCHE) Project 2.1 (Cardiovascular risks in early life and fetal echocardiography) and partly supported by the UKRI grant EP/X040186/1 (Turing AI Fellowship). Co-authors J. Alison Noble and Aris Papageorghiou were supported by the Oxford Partnership Comprehensive Biomedical Research Centre with funding from the NIHR Biomedical Research Centre (BRC) funding scheme.

\printcredits

\bibliographystyle{cas-model2-names}

\bibliography{myref}



\end{document}